\documentstyle[11pt,newpasp,twoside]{article}
\begin{document}
\title{AGB Variables as Distance Indicators}
\author{Michael Feast}
\affil{Astronomy Department, University of Cape Town, Rondebosch, 7701, South
Africa.\\ 
mwf@artemisia.ast.uct.ac.za}
\begin{abstract}
Evidence for the existence of Mira Period-Luminosity relations is
reviewed including recent work on Miras with thick circumstellar shells.
The calibration of the relation at $K$ is discussed in detail.
The nature of variables deviating from the PL relation and possible
metallicity effects are also discussed.
\end{abstract}
\section{Introduction}
   The possibility that at least a subset of AGB variables could
be used as distance indicators can probably be traced back to
the work of Gerasimovic (1928). He suggested on the basis of
statistical parallaxes 
that Mira variables showed a period -
(visual) luminosity relation. This was confirmed and extended in later
work (e.g. Gyllenberg 1929, 1930; Wilson \& Merrill 1942; Osvald \&
Risley 1961; Clayton \& Feast 1969). However, the scatter in visual
absolute magnitude at a given period, though difficult to estimate,
seemed quite large (perhaps $\sim$ 0.5 mag). The statistical parallax results
were extended to the near infrared and to $M_{bol}$ by Robertson \&
Feast (1981) on the basis of extensive $JHKL$ observations by
Catchpole et al.\,(1979) of nearby Miras. 
\section{The Slope of the Mira PL Relation}
\subsection{The Large Magellanic Cloud}
Glass \& Lloyd Evans (1981) and Glass \& Feast (1982)
found that a rather precise period-luminosity relation 
($\sigma \sim 0.25$mag)
existed at $JHK$
and $m_{bol}$ on the basis of 11 Miras in the LMC; these relations were
further refined by Feast (1984) to include the data of Wood, Bessell
\& Paltoglou (1985). This early LMC work depended on only a few (often
only one) infrared observations per star. 

An extensive programme
(Glass et al.\,1990) then allowed time-averaged
$JHK$ and bolometric  magnitudes to be derived for 29 O-rich and
20 C-rich Miras in the LMC (Feast et al.\,1989). At $K$ the PL relation
for O-Miras was found to be,
\begin{equation}
M_{K} = -3.47 (\pm 0.19) \log P + \beta
\end{equation}
the standard deviation from this equation is only 0.13 mag. Even though
some of this dispersion must be observational, it was found that there
was some evidence for a period-luminosity-colour relation. Such
a relation is of course to be anticipated since one can hardly expect the
Mira instability strip to have no width. Recently Glass \&  Lloyd Evans (2003)
have rederived periods for 
some of these stars from MACHO data. Their results
show that the above relation is rather stable since their rederivation
for 26 stars yields a slope of $-3.52 \pm 0.21$ insignificantly different
from the above value. In fact the small change is due to the
fact that they leave out three stars. It could be argued that at least two
of these might well be included as O-Miras 
(leading to a slope of $-3.49$) since their spectral types (Wood et al.\,1985)
are K? and S. Note that the designation (M),
derived from the photometry, in Feast et al.\,(1989)
implies only a likely O-Mira (i.e. K,M or S). The PL relation at $K$ for LMC
C-Miras appears to be the same as for the O-Miras.

Feast et al.\,(1989) also found a PL($M_{bol}$) relation for LMC O-Miras
with a slope of $-3.00 \pm 0.24$.
The data of Glass \& Lloyd Evans (2003) gives a slope of $-3.06 \pm 0.26$
(26 stars) or $-3.03$ (28 stars).
In the case of $M_{bol}$  the slope for the C-Miras is distinctly  different
($-1.86 \pm 0.30$). 
However, it has never been entirely clear whether
this difference is fundamental or due, for instance, to the method
of deriving the bolometric magnitudes. This could introduce systematic
errors since the spectra of the O- and C-Miras are so different (see also 
below). A PL slope in $M_{bol}$ for C-Miras derived by
Groenewegen \& Whitelock (1996) using both LMC and Galactic C-Miras of
known distance has a slope approximately midway between those for
O- and C-Miras just given.

The above discussion refers to Miras with periods shorter than about
420 days. In the sample of Feast et al.\,(1989) the Miras of longer
period fell above a linear extrapolation of the PL
relations both at $M_{K}$ and $M_{bol}$. Hughes \& Wood (1990) chose to fit
a PL relation of steeper slope for Miras of periods greater than 400 days.
It was later suggested on the basis of approximate theoretical arguments 
(Willson 2000) that this apparent steeping of the PL 
at longer periods was to be expected. However, the subsets of LMC Miras
so far discussed were all established in optical surveys. They thus
are strongly biased to Miras with thin circumstellar shells. In our own
Galaxy it is known that Miras with thick circumstellar shells are
more frequent amongst the longer period stars. A series of papers
(see references in Whitelock et al.\,2003) described the identification 
of luminous AGB stars in the LMC
from IRAS and their observation at $JHKL$ and with ISO. A subset of these
stars are large amplitude (i.e. Mira) variables and these were recently
discussed by Whitelock et al.\,(2003). Extensive $JHKL$ photometry of 
these stars
(including data from Wood et al. 1992 and Wood 1998) allowed the
derivation of periods and, together with IRAS and ISO data, bolometric
magnitudes. Both O-rich and C-rich objects are present in this sample.
For the C-rich objects the periods extend up to 939 days and for the
O-rich variables up to 1393 days. Whitelock et al. discuss in detail the
estimation of the bolometric magnitudes of these stars which can be done
in various ways. Whilst there remains uncertainty about the precise
bolometric magnitudes of these stars it is clear that the bulk of those
with periods greater than 400 days fall below the steep PL relation of
Hughes \& Wood (1990) and that both the O- and C-rich objects lie near
an extrapolation of the PL($M_{bol}$) for O-Miras of shorter period. 
Whitelock et al. point out that the long period C-Miras in
Magellanic Cloud clusters (Nishida et al.\,2000) also fall on an extrapolation
of this relation.
Extensive mid-infrared work would be required to determine
how much of the scatter about the PL relation in the sample
of Whitelock et al. was intrinsic.   

It is known that there are low amplitude semi-regular variables above the 
Mira PL relation (see e.g. Whitelock (1986), Bedding \& Zijlstra (1998) 
Wood (2000)).
Such stars are believed to be evolving towards the Mira PL relation. 
The nature of the long-period Miras above the PL relation has not been
clear. However,
Whitelock (see Whitelock \& Feast 2000a 
and Whitelock et al.\,2003) has pointed out that some at least of these
Miras are known to have strong lithium lines in their spectra 
(Smith et al.\,1995)
and this shows them to be undergoing Hot-Bottom-Burning (HBB). She therefore
suggests that this is the distinguishing characteristic of these objects.

It will be clear from the above that considerable caution is required in using
Miras with periods greater than 400 days as distance indicators, especially
in optically selected samples. An example of the problems that can arise
is provided the discovery of a 641 day, large amplitude, variable in the 
Local Group Irregular galaxy IC1613 (Kurtev et al.\,2001). This lies about 
0.7mag
above the PL relation at $K$ (for the adopted distance of this galaxy). 
Suspicion as to its true nature is aroused by the fact that it has a very
early spectral type (M3) and is rather blue ($J-K \sim 1.14$) for its long
period. It is possibly a HBB star or alternatively a relatively faint
supergiant variable.
\subsection{Is the PL Slope Universal?}
  Wood (1995) showed that the PL($K$) relation for SMC Miras has closely
the same slope as that in the LMC. Miras in Galactic globular clusters
fit the PL($K$) slope from the LMC
(Feast, Whitelock \& Menzies 2002). In the SgrI field of the Galactic Bulge,
Glass et al. (1995) found that the Miras (all probably O-Miras) fitted
a PL($K$) relation of slope $-3.47 \pm 0.35$, i.e. identical to that in 
the LMC. In this case the scatter about the relation is much higher
than in the LMC (0.35 mag compared with 0.13 mag). This difference is
primarily due to the depth of the Bulge in the line-of-sight.

In our own Galaxy, Glass et al.\,(2001) derived periods and mean $K$ magnitudes
for 406 Miras in an area 24 $\times$ 24
arcmin square around the Galactic centre. There is large and
variable extinction in this direction. The general distribution
of the stars in the $ K - \log P$ plane (their fig 4) 
nevertheless suggests that most
of these objects would fit a PL relation with the LMC slope. However,
Oritz et al.\,(2002) used ISO and near IR data to derive bolometric 
magnitudes for
OH/IR Miras  
within half a degree of the Galactic Centre.  These 
stars with relatively thick circumstellar shells show a wide
range in luminosity at a given period. Oritz et al. find that whilst 
many of their 
stars lie on a PL($M_{bol}$) relation derived from the SgrI stars of
Glass et al.\,(1995), a considerable number lie below this relation.
Thus there appears to be evidence, at least at long period, for Miras
below the PL $(M_{bol})$ relation near the Galactic Centre.
The matter cannot be considered
completely settled yet in view of the considerable difficulty in determining
the mean bolometric magnitude of the variables and also in estimating
the interstellar extinction. It is worth noting that if such 
``subluminous" OH/IR Miras were present in the LMC they might well have 
been too faint for existing surveys.
 
The most striking recent work on the Mira PL slope is that of
Rejkuba (2003) who established a well defined PL($K$) relation from 
about 1000 Miras
in the 
Giant Elliptical Galaxy Cen A.
The
infrared colours of these stars strongly suggests that they are 
mainly O-Miras. 
Most of these Miras have $\log P$ between 
2.3 and 2.7. The slope in Cen A ($-3.37 \pm 0.11$) clearly agrees well 
with that in the
LMC, $-3.47 \pm 0.19$.
\section{The Mira PL($K$) Zero-Point}
 A realistic determination of the Mira PL zero-point from trigonometrical
parallaxes became possible following the Hipparcos astrometric mission.
The Hipparcos Mira parallaxes are not of sufficient accuracy to allow 
their individual use in a determination of both the slope and zero-point
of a PL relation. However, since the results summarized above suggest
that the PL($K$) slope found for the LMC is closely the same elsewhere,
we can reasonably adopt this slope as applicable to local Galactic Miras.
It is then possible to use the method of reduced parallaxes to obtain
a zero point. Whitelock \& Feast (2000b) used this method together with
extensive infrared photometry (Whitelock et al.\,2000) to derive $\beta$
in equation 1 above
for O-rich Galactic Miras.

In Whitelock et al.\,(2000) (see also Whitelock 2003) 
it was shown that the shorter period Miras
could be separated by their colours into two sequences, the ``short-period-red"
and ``short-period-blue" stars. These two groups are also distinguished by
their kinematics. 
The latter belong to the main Mira sequence.
Whitelock et al. found evidence from the parallaxes that the 
``short-period-red" Miras were brighter at a given period than those of the
main Mira sequence. Whilst the main Mira PL sequence is believed to 
represent the end points of
AGB evolution for stars of different initial masses and/or metallicities,
the evolutionary status of the ``short-period-red" stars is still
uncertain.

Whitelock \& Feast (2000b) found from Hipparcos parallaxes of O-Miras
(omitting ``short-period-red" stars) that $\beta = 0.84 \pm 0.14$ mag.  
This result depends on data for 180 Miras though most of the weight
resides in a relatively small number of stars.
It has 
recently been suggested that the Hipparcos results for some Miras
require a recalibration of the adopted corrections for chromaticity
(Platais et al.\,2003, Pourbaix et al.\,2003). Knapp et al.\,(2003) have
published revised parallaxes based on these new corrections. Using
these new parallaxes Whitelock 
(Leiden Workshop on the Future of AGB Star Research 2003 and to be published) 
finds a zero-point of $\beta = 1.04 \pm 0.13$ from 140 stars. The change in 
zero-point is partly due to the fact that the selection of stars is
slightly different since Knapp et al. do not give revised parallaxes for
all the Hipparcos Miras. A discussion of statistical bias corrections
(Feast 2002) suggests that a small correction of $\sim 0.02$mag is necessary
leading to $\beta = 1.06$.
Table 1 lists this value of the PL($K$) zero-point together with two other
Galactic determinations. Vlemmings et al.\,(2003) have obtained distances
to four Mira variables using VLBI astrometry of OH maser spots and
this leads to the value of $\beta$ listed. The final figure in the table
is from Miras in globular clusters with cluster distances based on
Hipparcos parallaxes of subdwarfs (Feast et al.\,2002).
\begin{table}
\caption{The Zero-Point of the Mira PL(K) Relation}
\begin{tabular}{cc}
Method & $\beta$\\
Parallaxes & 1.06 $\pm$ 0.13\\
OH  VLBI       & 1.01 $\pm$ 0.13\\
Globular Clusters & 0.93 $\pm$ 0.14\\
&\\
Mean & 1.00 $\pm$ 0.08\\
\end{tabular}
\end{table}

Whitelock \& Feast (2000) also analysed the parallaxes of Galactic C-Miras.
However, in view of the uncertainties, which they discuss, it would
appear best at present to rely on the LMC result that the 
(unobscured) C- and O-Miras
fit the same PL($K$) relation (Feast et al.\,1989). It has already
been noted (see Whitelock et al.\,2003) that 
within the uncertainties both the O- and C-rich Miras
fit, the
same $M_{bol}$-$\log P$ relation at least at long periods.

\section{Is the PL($K$) Zero-Point the Same Everywhere?}
 There are a number of tests that can be applied to see whether the
above PL($K$) calibration is applicable to all O-Miras.

(1) As Table 1 shows the value of $\beta$ from Galactic field Miras and
those in Globular Clusters is not significantly different.

(2) The mean value of $\beta$ from Table 1 (1.00) when used with the
results of Feast et al.\,(1989) leads to an LMC modulus of
$18.48$ with an uncertainty of about 0.1 mag or slightly less. This 
agrees with an LMC modulus based on Cepheids calibrated in a variety 
of manners (Feast 2003) which is 18.52 and has a similar small
uncertainty.

(3) The difference between the moduli of the SMC and LMC derived from
the Mira PL($K$) relation is $\sim 0.4$mag (Wood 1995) which compares
well with the difference derived from observations of the magnitudes
at the red giant tip (0.44 mag, Cioni et al.\,2000). A similar result
is obtained from Miras in $M_{bol}$ (Cioni et al.\,2003).

(4) The results of Glass et al.\,(1995) on Miras in the SgrI field of the 
Galactic Bulge together with the calibration of Table 1 yields
a distance to the Galactic centre ($R_{0}$) of $8.5 \pm 0.7$kpc. This may be 
compared with $8.5 \pm 0.5$kpc derived from Cepheid kinematics 
(Feast \& Whitelock 1997) and $8.0 \pm 0.4$kpc recently derived from the
orbit of a star very close to the central black hole (Eisenhauser et al.\,2003).

(5) With the Galactic calibration of Table 1, the Mira PL($K$) results
of Rejkuba (2003) for Cen A yield a distance modulus for this
giant elliptical of $28.0 \pm 0.2$mag agreeing with the RGB tip distance
derived by Rejkuba, $27.9 \pm 0.2$ mag.

These tests indicate that there is no present evidence for any variation
of the PL($K$) zero-point at least within the observational uncertainties.

\section {Are there Metallicity Effects on the PL relation?}
 It is known that at least in globular clusters there is a
relation between [Fe/H] and Mira period (see Feast \& Whitelock 2000).
We then need to know whether there are variations in
chemical abundance at a given period between Miras in different systems.
That such seems likely between our Galaxy and the LMC was shown by
Glass et al.\,(1995). They found that the $JHK$ colour-period relations
were different in the SgrI field of the Galactic Bulge from those in the 
LMC. This question has been discussed by Feast (1996) and Feast \& Whitelock
(2000). The Galaxy-LMC difference can be attributed to the 
weakening of the infrared $\rm H_{2}O$ band due to a relative underabundance
of oxygen or an overabundance of carbon (locking up oxygen in CO) in the LMC.
Whether this also implies an overall metal deficiency at a given period
in LMC Miras compared with those in our Galaxy is not certain since
the [O/Fe] ratio for these stars is not known. 
 Whilst the empirical results discussed above indicate that the PL($K$)
relation holds in all the places to which it has so far been applied,
the colour differences between the SgrI field and the LMC show that
PL relations at  some other colours may not be universal. 
For instance Feast \&
Whitelock (1999) point out that that there is a difference in the
$(K - m_{bol})$-$\log P$ relations in the LMC and the SgrI field.
This suggests that a PL($M_{bol}$) relation is more metallicity sensitive
than a PL($K$) relation, as proposed theoretically by Wood (1990).
\section{Can Other AGB Variables be used as Distance Indicators?}
 Wood (2000) showed that AGB variables in the MACHO survey of the LMC
defined a number of sequences in the $K-\log P$ plane. Similar
sequences are found in the Galactic Bulge (Glass \& Schultheis 2003) 
These sequences show considerable internal scatter. Some of this
will be due to the fact that the 
results depend on only single observations at $K$. However since only
the Miras (in Wood's sequence C) are expected to have large $K$ amplitudes
(most of the stars are small amplitude semiregular variables) the
scatter in the LMC must (except for the Miras) be mainly 
intrinsic\footnote{In the Bulge there is addition scatter due to the depth 
in the line of sight.}.
It remains possible that such sequences might be
narrowed by the introduction of a colour term. Even so, it is not at
present generally possible to determine to which of Wood's sequences a field
SR variable belongs. Thus the use of these sequences for distance
determination seems limited for individual stars. 
\section{Conclusions}
 So far as can presently be determined the slope and zero-point of the
Mira PL relation at $K$ is invariant. The rather precise zero-point of this
relation together with the small scatter of individual stars about the
relation now makes it possible to derive good distance for Miras
in our Galaxy and in the Local Group. The spectacular work of Rejkuba
on Miras in Cen A indicates the potential for Mira studies well
beyond the Local Group.  
 Amongst important problems still to solve are; the evolutionary status of
the hot bottom burning Miras above the PL relation; the nature of the 
Miras which apparently lie below the PL$(M_{bol})$ relation in the
region of the Galactic Centre; and, the nature of the Galactic short-period-red
Miras discussed in section 3. In addition there 
is the problem of understanding the small percentage of Miras
which show
long term period changes. These have usually been attributed to shell-flashing
episodes (Wood \& Zarro 1981) but it has recently been suggested as an
alternative that they are due to epochs of envelope relaxation
(Zijlstra, Bedding \& Mattei 2002).
\section{Acknowledgements}
I am grateful to Patricia Whitelock for her help and advice and to
Marina Rejkuba for information in advance of publication.

\end{document}